# CdSe-single-nanoparticle based active tips for near-field optical microscopy

N. Chevalier[†], M. J. Nasse[†], J .C. Woehl[†§], P. Reiss [‡], J. Bleuse[‡], F. Chandezon[‡] and S. Huant[†]*

[†] Laboratoire de Spectrométrie Physique
CNRS and Université Joseph Fourier Grenoble
BP 87, 38402 Saint Martin d'Hères, France

[‡] CEA Grenoble, Département de la Recherche Fondamentale sur la Matière Condensée
17, rue des Martyrs, 38054 Grenoble Cedex 9, France

*Corresponding author. Email: Serge.Huant@ujf-grenoble.fr

[§] Present address: Department of Chemistry and Biochemistry, University of Wisconsin, Milwaukee, 3210 N. Cramer Street, Milwaukee, WI 53211, USA

**Abstract**

We present a method to realize active optical tips for use in near-field optics that can operate at room temperature. A metal-coated optical tip is covered with a thin polymer layer stained with CdSe nanocrystals or nanorods at low density. The time analysis of the emission rate and emission spectra of the active tips reveal that a very small number of particles - possibly down to only one - can be made active at the tip apex. This opens the way to near-field optics with a single inorganic nanoparticle as a light source.

**Short title**

Active optical tips for near-field microscopy.

**PACS numbers:**

Current developments in aperture Near-Field Scanning Optical Microscopy (NSOM) include the search for improved spatial resolution in the nanometer range where a novel optics is expected to emerge. Active tips made of a fluorescent nano-object attached to a regular tip [1] have much to offer in this context. Apart from challenging the ultimate resolution in optics, there are more fundamental *stimuli* to this research. For instance, standard electric-dipole selection rules in semiconductor quantum dots are expected to be strongly affected by multipolar contributions at ultra-high spatial resolution [2]. This is especially true when the latter approaches typical length scales of the semiconductor material such as its exciton Bohr radius [3]. In addition, a point-like source of light attached to a NSOM aperture tip has been argued to be the ideal tool for mapping all details of the photonic Local Density-of-States (LDOS), which is a fundamental near-field property of a surface [4].

A major breakthrough towards nanooptics with active tips has been realized by Michaelis *et al.* [1] who have performed NSOM imaging with a single molecule serving as a point-like source of light that operates at low temperature. This pioneering work has been extended to other systems like color centers in a diamond microcrystal [5] and rare-earth doped glass micro-particles [6]. Recently, Shubeita *et al.* [7] have made a further step in the direction of active tips providing high spatial resolution. A thin polymer layer with a large number (of the order of 1000) of CdSe nanocrystals (NCs) has been deposited on coated optical tips. These active tips have been used at 300 K for NSOM Fluorescence Resonance Energy Transfer (FRET) imaging of single dye molecules. An optical resolution of 90 nm has been achieved. However, none of the works published so far is free from limitation such as the low-temperature operation or/and the limited spatial resolution due to the size of the supporting objects or number of active particles involved.

In this letter, we describe the emission properties of active tips made of a few CdSe nanoparticles (NPs), either NCs or nanorods (NRs), deposited at the apex of a tapered and coated tip. Our goal is to make photostable light nanosources working at room temperature and potentially able to offer optical resolution in the sub-10 nm range. Semiconductor NPs have a large potential in this respect because they have natural small dimensions, their optical properties are size-tunable, they possess an absorption threshold - rather than absorption bands as dye molecules – they are optically active at room temperature, and they are more photostable (little bleaching) than single molecules [8].

ZnSe capped CdSe NCs have been synthesized following the procedure of Reiss *et al.* [9]. The used core NCs have a diameter of 3.8 nm. After growing a ZnSe shell of approximately 1 nm thickness on their surface, the resulting CdSe/ZnSe core/shell NCs exhibit a room-temperature photoluminescence (PL) quantum yield of *ca*. 70% in chloroform and a size dispersion < 10 %. With the goal to improve their stability against photo-oxidation, the surface of the CdSe/ZnSe NCs has been functionalized with tridecanedithioic acid [10]. Recently, several advances in the synthesis of colloidal semiconductor NCs have been made, allowing for size and shape control [11,12]. Of particular interest in this context is the possibility to obtain quantum confined CdSe NRs with a narrow distribution of length and diameter. The NRs [12] used in this work have a diameter of 3.9 nm and a length of 13 nm and are coated with a ZnSe/ZnS double shell. The intermediate ZnSe shell provides strain relaxation at the interface between the CdSe core and the outer ZnS shell [13], in a similar manner as in the CdSe/CdS/ZnS system [14]. After shell growth, the NRs have a diameter of 4.4 nm and a length of 21.5 nm (aspect ratio 4.9).

Before depositing a small number of CdSe NPs at the optical tip apex, it is necessary to characterize their emission properties in a more conventional environment. This is done at room temperature (all measurements presented in this paper are performed at room temperature) by means of scanning confocal microscopy of a polymer (PMMA) film containing CdSe NPs. Samples are prepared by spin coating the film onto a glass cover slip. CdSe NPs are excited with the 458 nm spectral line of a cw $Ar^+$ laser through a X100 infinity-corrected objective (NA=1.3). The PL signal, which is selected in the spectral window 540-620 nm with suitable optical filtering, is collected via the same objective.

A first series of measurements consists in detecting this PL signal by an avalanche photodiode in photon counting mode. PL images are then acquired by raster scanning the sample through the excitation spot of the microscope. Four fluorescence images of CdSe-NPs doped PMMA film are shown in Figure 1. The typical blinking behavior [15,16] of several isolated objects can be clearly observed both for NCs and NRs. The "off" periods of the PL emission can be seen as horizontal dark stripes in some of the PL spots. As a consequence of this blinking, the images are only partly acquired. Nevertheless, up to four individual NCs can be distinguished in Figure 1a and Figure 1b, with possibly a very weak fifth one in the lower left corner, whereas presumably only one NR is seen on each image of Figure 1c and Figure 1d. Although not elucidated so far [17], the PL intermittency phenomenon is commonly attributed to electron (or hole) ejection via an Auger process [15], which leaves a charged particle switching the nanocrystal in a "dark" state: other photogenerated electron-hole pairs will recombine non-radiatively until the excess charge leaves the nanocrystal.

In Figure 2a, respectively Figure 2b, the PL intensity is plotted as a function of time for NCs and NRs, respectively. Here, scanning has been stopped, in such a way that the laser spot is focused on a single NP previously isolated in a PL image similar to those of Figure 1. It is clearly seen that the single NPs alternate "on" and "off" periods with some "off" having duration in excess of 100 s. Although this blinking can raise problems in some applications, it turns out to be an advantage here because it helps us to evaluate the number of active NPs at the apex of the optical tip as described later in the paper. Indeed, isolated particles blink in a discrete manner as shown in Figure 1, Figure 2a and Figure 2b whereas a large ensemble of NCs does not exhibit such a behavior. This is because individuals of a large ensemble are not synchronized and an average behavior is revealed.

A second characterization consists in spectrally analyzing the filtered PL signal by means of a spectrometer equipped with a cooled charged-coupled device. Figure 2c, respectively Figure 2d, compares spectra of a single CdSe NC, respectively NR, with spectra of the corresponding ensemble. In both cases, single NPs have gaussian spectra that are substantially narrower - typically 20 nm in the wavelength domain - than the ensemble. The full width at half-maximum (fwhm) of the ensemble is 35 nm for NCs and even 40 nm for NRs, respectively. This reflects the size dispersion of NPs. In principle, the blinking behavior of single NPs should also manifest itself in the spectra (see below), but the large integration time of 5 minutes required to have sizable signal strength in our confocal spectrometer hindered such an observation. The behavior observed here in both the PL time traces (i.e. discrete blinking) and PL spectra (rather narrow spectra) of single NPs serves as a valuable reference for characterizing them when located at the apex of an optical tip as described in the next section.

Since the fluorescence signals emitted by a few NPs attached to an optical tip are expected to have very low intensity, it is most important to start with substrate tips delivering an extremely low parasitic fluorescence signal. Our tips have been prepared by a special chemical-etching process of pure silica optical fibers and they have been subsequently coated with aluminum. They offer a free optical aperture at the apex in the 150 nm range and a large transmission of the order of $10^{-2}$. Details of their fabrication procedure and of their optical properties will be given elsewhere. CdSe single-NP-based active optical tips are prepared in the same manner as Shubeita *et al.* [7] but with a very small number of CdSe NPs on the active area of the tip. Initial solutions of CdSe/ZnSe NCs (respectively NRs) in chloroform with a concentration of 5 mg/ml (respectively 10 mg/mL) are used. These solutions are manifold diluted (2500 for NCs, 12500 for NPs) and are then mixed with a PMMA polymer in order to obtain 2 % in-weight PMMA solutions. The metal-coated fiber probe is briefly dipped into, and then withdrawn from, one of these solutions. After evaporation of the solvent, a thin PMMA layer stained with CdSe NPs is formed at the tip apex (Figure 3). An analysis by scanning electron microscopy confirms that the thickness of the layer is well below 100 nm although it is not possible to give a more quantitative figure at this stage. The laser beam at 458 nm is coupled to the cleaved end of the fiber probe and the PL signal is collected with a microscope objective (X60, NA= 0,75). Appropriate optical filters are arranged in front of the detector to restrict the relevant spectral range to 540-620 nm. As for the test samples, the PL emitted by the active probe is either analyzed through its time evolution or spectrally dispersed. A scheme of the experimental set-up is given on Figure 3.

The four successive time traces shown in Figure 4a, respectively Figure 4b, reveal a clear switching between "on-like" and "off-like" intensities for NCs, respectively NRs. This mimics the behavior of single NPs in a PMMA film described above. This is a strong indication that the number of active NPs involved is certainly of the order of a few units only. This number goes down to zero over some periods of time since the "off" intensity matches the small background fluorescence of the substrate tip. This behavior is very clear for NCs (Figure 4a) but is less marked for the optical tip stained with NRs (Figure 4b) which showed an unusual strong background fluorescence already prior to depositing the thin PMMA layer. This is possibly due to undesired pollution at the tip apex. Therefore, in the rest of the letter, we will discuss in more detail the CdSe-NCs based tip only, although NRs give very similar results.

The trend, suggested by the time traces of Figure 4a, that only a few NCs contribute to the PL signal is confirmed by the spectral analysis of Figure 5a which shows in a (time,wavelength) plane 42 successive spectra of the PL emitted by the active tip. Each spectrum has been collected over 30 s, fitted by a multi-peak gaussian curve, and normalized. The PL intensity is represented with a stretched false color in the (time, wavelength) plane. Clearly, several periods without PL emission at all (red) can be seen, e.g., at around 200 s, 400 s, and 1200 s along the time axis. This corresponds to an "off" state extending at least over 30 s. In addition, it can be seen that the emission wavelength(s) change(s) dramatically with time during the "on" periods. This manifests itself in Figure 5a as large shifts in peak positions (purple) along the wavelength axis. This is due to a combination of a change in time of the number of NCs being actually active and of possible spectral diffusions of individual NCs [18]. An interesting subsidiary finding in our experiments is that not all of the CdSe nano-emitters are quenched at the tip apex despite the proximity of the aluminum coating, see Figure 3.

Figure 5b is a selection of a few spectra taken from the above series that confirms that the PL spectra change dramatically with time. The spectrum at the middle (bottom) shows two (one) peak(s) with a fwhm of the order of 15 to 20 nm similar to that shown in Figure 2c for a single NC. The upper spectrum does not show any PL at all which suggests that all of the NCs have turned off during the collection of this spectrum, in agreement with the red lines seen in Figure 5a. Again this is reminiscent of the blinking behavior observed for single NCs dispersed in a PMMA film. Due to the limited integration time of 30 s, it is indeed possible here to observe the blinking in the spectral domain in contrast with Figure 2c (integration time ten times longer).

As a complementary test, similar measurements have been done on an active tip stained with a high concentration of CdSe/ZnSe NCs. In Figure 6, no blinking is observed neither in time traces nor in spectra. These observations give strong indication that only a few NCs contribute to the PL of the active tip in Figure 4a and Figure 5.

In summary, a method for elaborating CdSe NPs-based active NSOM tips operating at room temperature has been described. Both the spectral and temporal behavior of the active tips confirm that only a limited number of NPs, either NCs or NRs, are actually active at the tip apex. This opens the way to near-field optics using a single semiconductor NP as light source - a nanosystem potentially much more photostable than a single molecule. An obvious on-going extension of the present work is to perform NSOM imaging of test surfaces using these active tips.

ACKNOWLEDGMENT   The substrate optical tips used in this work have been prepared by J.F. Motte. We are thankful to C. Querner for the preparation of tridecanedithioic acid and M. Stark for stimulating discussions. Financial supports from Institut de Physique de la Matière Condensée (IPMC) Grenoble (SoLuNa project) and from Action Concertée Nanosciences 2004 (NANOPTIP project) are gratefully acknowledged.

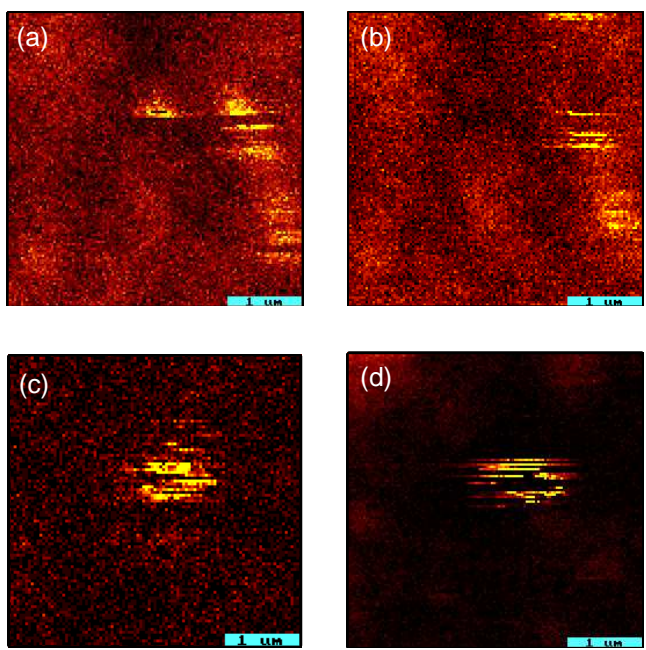

**Figure 1.** Scanning confocal PL images of isolated NPs. (a-b) CdSe/ZnSe NCs. Here, two successive images of the same region have been recorded. The scanning area is 4×4 µm$^2$ and the integration time is 4 ms/pixel. The excitation density is approximately 90 W/cm$^2$. (c-d) CdSe/ZnS NRs. Here, images are taken for two different regions of the sample. The scanning area is 4×4 µm$^2$ and the integration time is 3 ms/pixel. The excitation density is approximately 400 W/cm$^2$.

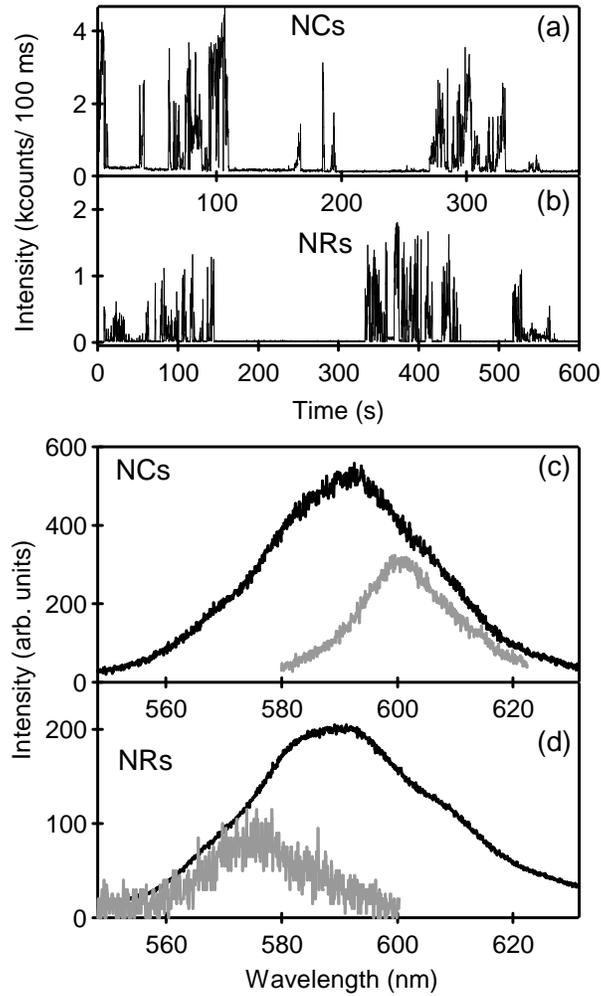

**Figure 2.** (a-b) PL time traces of single CdSe NPs embedded in a PMMA film (binning time = 100 ms). (a) is for a single CdSe/ZnSe NC. The excitation intensity is 90 W/cm$^2$. (b) is for a single CdSe/ZnS NR. The excitation intensity is 400 W/cm$^2$. (c-d) PL spectra of single NPs compared with the ensemble. (c) is for CdSe/ZnSe NCs and (d) for CdSe/ZnS NRs. For the ensembles (black lines), the integration time is 1 s only, whereas for single NPs (grey lines), the integration time is 5 min.

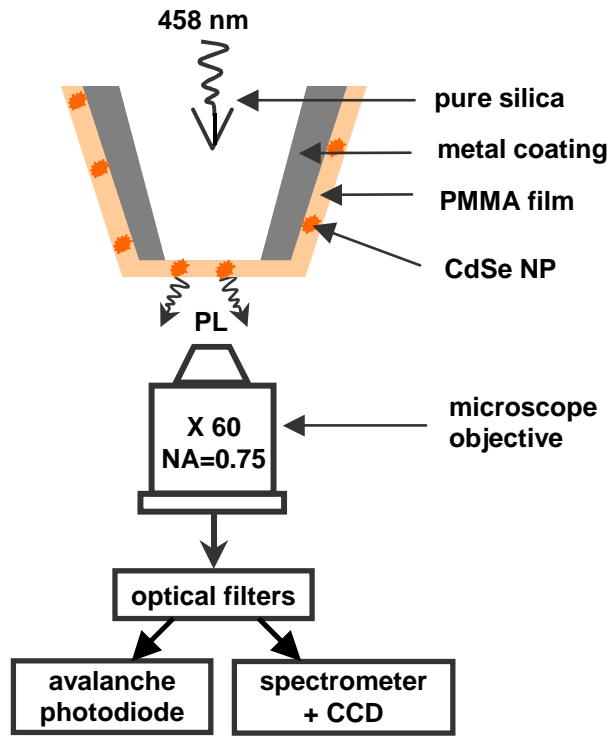

**Figure 3.** A scheme of an active tip and of the optical set-up used for characterizing its emission properties both in the time and in the spectral domains.

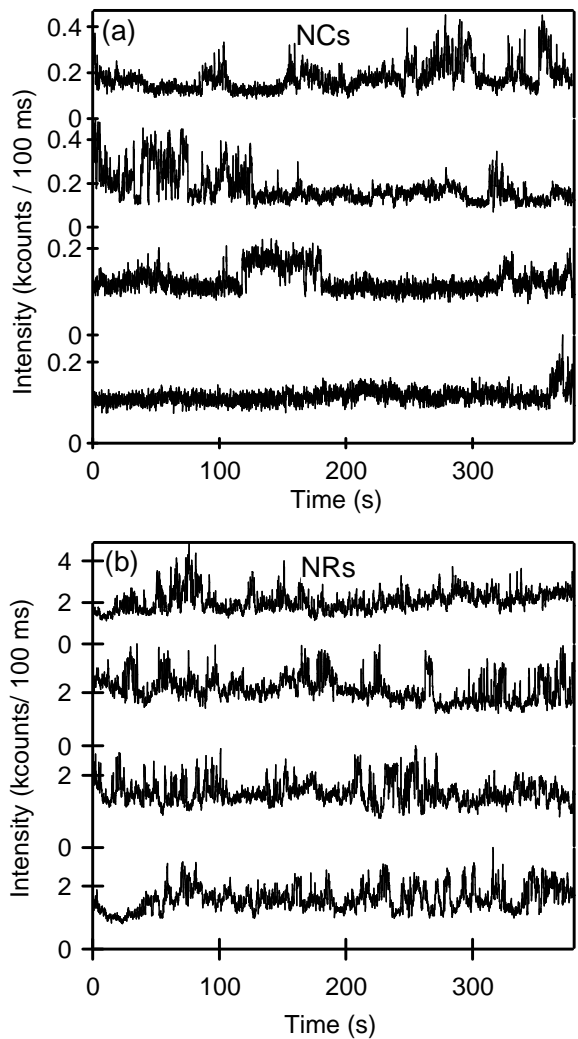

**Figure 4.** Four successive PL time traces of CdSe NPs embedded in a PMMA film at the apex of an optical tip (binning time = 100ms). (a) CdSe/ZnSe NCs, the excitation intensity is 230 W/cm$^2$. (b) CdSe/ZnSe NRs, the excitation intensity is 170 W/cm$^2$.

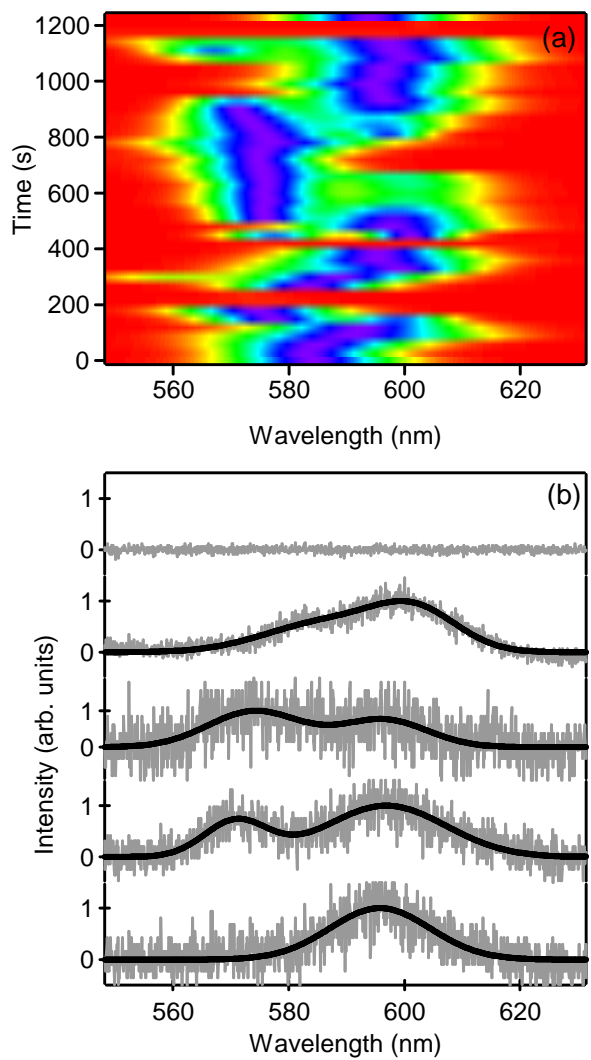

**Figure 5.** (a) Spectral analysis of CdSe/ZnSe NCs on the active probe. The intensity is normalized and is represented with stretched false colors. Red corresponds to vanishingly small PL intensity, purple to high intensity. A series of 42 successive spectra is shown as a function of time. Each spectrum is collected over 30 s. The excitation intensity is 230 W/cm$^2$. (b) A few experimental spectra selected from the series shown in (a). Also shown is multi-peak fitting of the spectra.

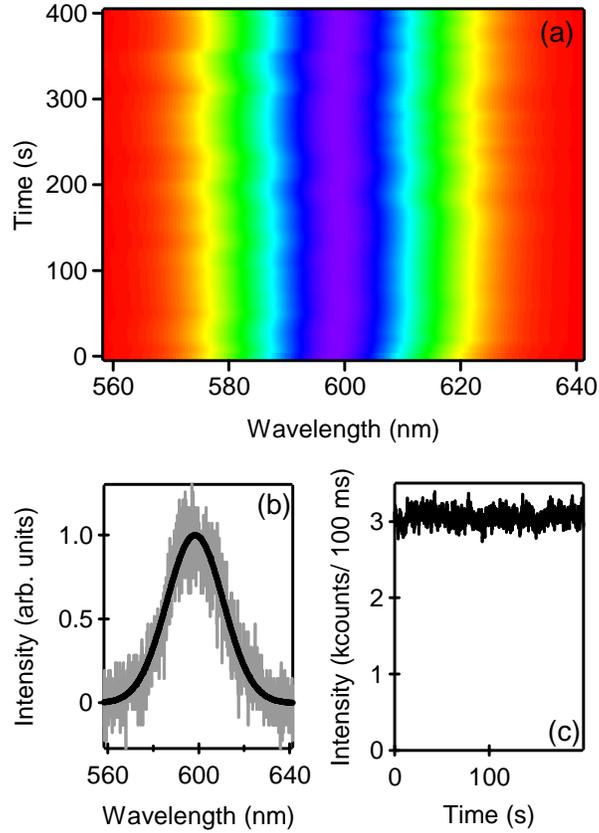

**Figure 6.** (a) Spectral analysis on an active tip stained with a high concentration of CdSe NCs. The intensity is normalized and is represented with stretched false colors similar to Figure 5. The integration time is 10 s, the excitation intensity is 200 W/cm$^2$. (b) A typical PL spectrum extracted from the series in (a). (c) PL time traces of an active tip stained with a high concentration of NCs. The excitation intensity is 4 W/cm$^2$ (binning time = 100 ms).